# Technical and Economic Feasibility Analysis of Underground Hydrogen Storage: A Case Study in Intermountain-West Region USA

Fangxuan Chen[1,2,*], Zhiwei Ma[1], Hadi Nasrabadi[2], Bailian Chen[1], Mohamed Mehana[1,*], Jolante Wieke Van Wijk[1]

1) Los Alamos National Laboratory, Los Alamos, United States
2) Texas A&M University, College station, United States
*Corresponding author: Fangxuan Chen: fangxuanche@lanl.gov, Mohamed Mehana: mzm@lanl.gov

**Abstract**
Hydrogen is an integral component of the current energy transition roadmap to decarbonize the economy and create an environmentally-sustainable future. However, surface storage options (e.g., tanks) do not provide the required capacity or durability to deploy a regional or nationwide hydrogen economy. In this study, we have analyzed the techno-economic feasibility of the geologic storage of hydrogen in depleted gas reservoirs, salt caverns, and aquifers in the Intermountain-West (I-WEST) region. We have identified the most favorable candidate sites for hydrogen storage and estimated the volumetric storage capacity. Our results show that the geologic storage of hydrogen can provide at least 72% of total energy consumption of I-WEST region in 2020. We also calculated the capital and levelized costs of each storage option. We found that a depleted gas reservoir is the most cost-effective candidate among the three geologic storage options. Interestingly, the cushion gas type and volume play a significant role in the storage cost when we consider hydrogen storage in saline aquifers. The levelized costs of hydrogen storage in depleted gas reservoirs, salt caverns, and saline aquifers with large-scale storage capacity are approximately $1.3, $2.3, and $3.4 per kg of $H_2$, respectively. This work provides essential guidance for the geologic hydrogen storage in the I-WEST region.

## 1. Introduction

Greenhouse gas emission is a major cause of climate change, which has largely affected the earth's ecology and environment (Qiu et al. 2020; Mouli-Castillo et al. 2021). It is estimated that fossil fuel combustion leads to 74% of total greenhouse gas emissions (Mouli-Castillo et al. 2021). Therefore, cleaner energy alternatives are utilized to reduce carbon emissions, including solar, wind, hydropower, bioenergy, and geothermal energy (Ellabban et al. 2014). However, renewable energy sources are often seasonal and/or location-dependent and cannot provide constant and reliable energy to meet the energy requirements. To solve this problem, excess energy should be stored for future use. $H_2$ serves as a clean energy carrier, which can be stored both into surface tanks and into subsurface sites geologically. **Figure 1** compares the current $H_2$ storage options in terms of discharge duration and power (Ye et al. 2022). It is of great significance to assess the feasibility of $H_2$ geolgic storage.

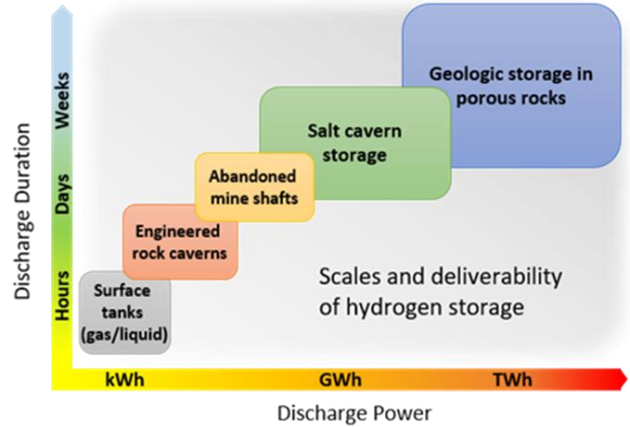

**Figure 1**. Scales and deliverability of hydrogen storage.

Recently, the feasibility assessment of $H_2$ geologic storage has drawn the attention of various research institutes around the world (Le Duigou et al. 2017; Stone et al. 2009; Liu et al. 2020). Scafidi et al. (2021) quantified the $H_2$ storage capacity of gas fields and saline aquifers on the UK continental shelf. They estimated that, assuming 50% of cushion gas, the working gas capacities were 6900 TWh and 2200 TWh for gas fields and saline aquifers, respectively. Liu et al. (2020) investigated the feasibility of $H_2$ storage in salt caverns in Jiangsu province, China. They used numerical simulation models to evaluate salt cavities' stability and tightness. The simulation results demonstrated that the gas permeability of interlayers should be less than $10^{-3}$ mD to ensure wall tightness. Jahanbani Veshareh et al. (2022) investigated the chemical and biochemical reactions in chalk hydrocarbon reservoirs to assess their feasibility as $H_2$ storage sites. Danish North Sea chalk hydrocarbon reservoirs were selected as targeted sites, and four principle reactions (e.g., abiotic calcite dissolution and biological souring) were considered. Their results proved that chalk reservoirs were not affected by the chemical or biochemical risks and are good candidates for $H_2$ storage. Zeng et al. (2022) analyzed the effects of carbonate dissolution on $H_2$ loss for the $H_2$ storage in carbonate reservoirs. Their results suggested that 6.5% of $H_2$ would be consumed for six-month storage. In addition, the $H_2$-brine-rock interactions would generate a large amount of methane, leading to reduced $H_2$ purity.

The conventional $H_2$ storage options include salt caverns, saline aquifers, and depleted gas reservoirs (Muhammed et al. 2022). Different storage sites have different characteristics and can be used for various purposes. For instance, salt caverns are created by solution mining in salt-rich formations (Lemieux et al. 2019). Based on the geologic structure, two types of salt caverns can be utizlied: domal salt caverns and bedded salt caverns. A domal salt cavern has an integrated cavity created in a thick rock salt layer. In bedded salt caverns where rock salt layers are discontinuous, the cavity is built only in rock salt layers, which leads to a disconnected cavity (Bruno 2005), **Figure 2**. Salt caverns have the advantages of long-term structural stability, good seal integrity, and low cushion gas requirement (Ramesh Kumar et al. 2021; Wallace et al. 2021). In addition, the high salinity nature strongly restrains the microbiological activities, which are detrimental to the storage operations (Dopffel et al. 2021).

However, salt cavern construction requires a large amount of water during the leaching process, which might be challenging in water scarcity regions. Consequently, water injection and disposal are needed during the cavern creation stage (Zivar et al. 2021). It is worth noting that several salt caverns have been built in the United States and the United Kingdom (Tarkowski 2019).

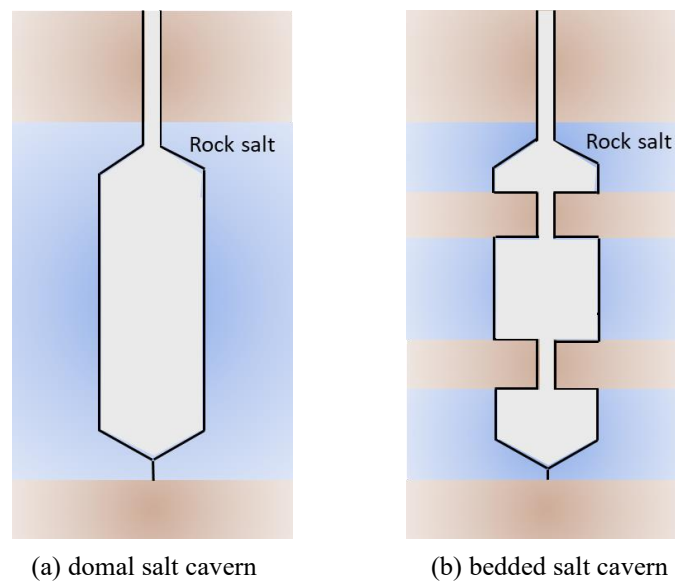

(a) domal salt cavern  (b) bedded salt cavern

**Figure 2**. Schematic figure of (a) domal salt cavern and (b) bedded salt cavern. The blue region represents the rock salt layer. The white region is the salt cavern. The domal salt cavern has an integrated cavity created in a thick rock salt layer, while bedded salt cavern has a compartmental cavity due to the discontinuous rock salt layers.

Also depleted gas reservoirs are favorable storage sites for various gases. As previous geologic hydrocarbon traps, these reservoirs have a large-scale porous media and impermeable seal (Tarkowski 2019; Singh 2022). In addition, the geologic characteristics of these reservoirs have been described in detail, and existing infrastructures reduce the initial capital investment. Moreover, the residual gas in the reservoir can serve as the cushion gas, reducing the amount of cushion gas required. However, residual gas also might impact the purity of $H_2$ during the extraction process.

Aquifers are abundant in sedimentary basins and can also be used for $H_2$ storage if salt caverns and depleted gas reservoirs are unavailable in the region. Generally, an ideal aquifer for $H_2$ storage should have two characteristics: 1) water-bearing sand with high porosity and permeability; 2) both vertical and lateral seals. An impermeable cap rock with an anticline shape is preferred because it helps form a gas cap, reducing the amount of water produced during $H_2$ extraction (Katz and Tek 1970). Apart from $H_2$ leakage potential, the potential $H_2$ reactivities with saline water should be carefully assessed. Currently, no pure $H_2$ storage in an aquifer is reported worldwide (Sainz-Garcia et al. 2017). However, reservoir simulations have been conducted to investigate the feasibility of $H_2$ storage in the saline aquifers (Azretovna et al. 2020).

Cost analysis is an important aspect of large-scale $H_2$ geologic storage (Gorre et al. 2020; Blanco et al. 2018). Taylor et al. (1986) divided the total cost of $H_2$ storage into three parts: capital cost, operating cost, and additional investment. The capital cost includes the storage site development, equipment purchase, general working system (heating, lighting, monitoring, and alarm system), well, and surface pipeline network. The operating cost includes $H_2$ production, water cooling, power, and labor costs. The additional investment involves the costs of land usage and plant construction. Ugarte and Salehi (2021) mentioned that the materials used in $H_2$ storage should be resistant to corrosion and rusting, which leads to an extra embrittlement cost. Lord et al. (2014) analyzed the total capital cost and levelized cost of $H_2$ storage in salt caverns, depleted oil and gas reservoirs, hard rock, and aquifers. Their results showed that depleted oil & gas reservoirs have the lowest levelized cost of $H_2$ storage, which are the most economical storage candidates. The significant cost of $H_2$ storage in salt caverns and hard rock is from the mining cost, while the cost of cushion gas accounts for most of the expenditure in depleted gas reservoirs and saline aquifers.

The I-WEST region consists of Arizona, Colorado, New Mexico, Montana, Utah, and Wyoming, which account for 6% of the population and 18% of the total area of the United States (2022 World Population by Country). The six states share similar energy challenges: water scarcity and economic dependency on fossil fuels. The net $CO_2$ emissions from fossil fuels in the I-WEST region in 2019 are shown in **Figure 3**. To reduce the carbon emissions and alleviate the dependency on fossil fuels, various initiatives and projects have been started to design an energy transition roadmap. Such a roadmap includes $CO_2$ capture and storage, $H_2$ production, storage and transportation, biomass utilization, and conversion components. Due to the legacy of the oil and gas industry in the region, many depleted gas fields are potential geological sites for $H_2$ storage. In addition, the widely deposited rock salts are ideal places for salt caverns of hydrogen storage.

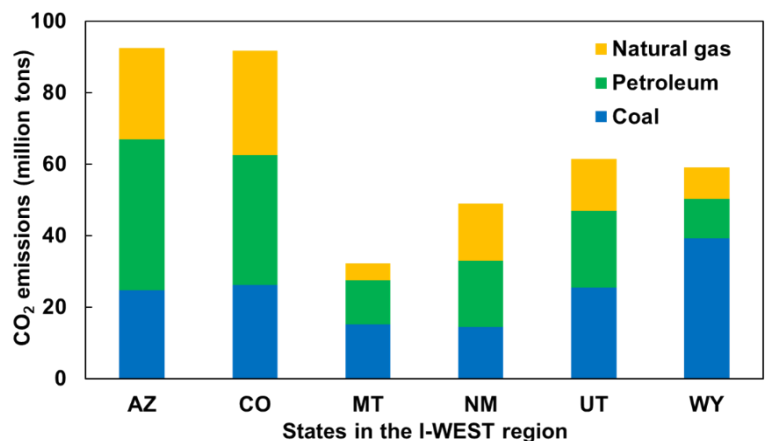

**Figure 3**. $CO_2$ emissions in I-WEST region in 2019. AZ: Arizona, CO: Colorado, MT: Montana, NM: New Mexico, UT: Utah, WY: Wyoming. The data were obtained from Energy Information Administration (EIA 2022).

Energy transition in the I-WEST region entails wide adoption of clean energy alternatives to replace conventional fossil fuels (McPherson et al. 2018). With the rapid development of $H_2$ production techniques, the cost of $H_2$ production has been largely reduced, which makes it a good alternative to fossil fuels (Reuß et al. 2017; Gorre et al. 2019). As a result, H₂ storage becomes an urgent challenge that needs to be addressed.

We analyzed the H₂ storage capacity in potential geologic sites and estimated the cost of H₂ storage of different types of geologic sites in the I-WEST region. The ideal candidates for H₂ storage sites are identified, together with the H₂ storage capacity of each site. The capital and levelized costs of H₂ storage in three specific geological sites are estimated. In addition, the effects of storage volume and cushion gas type on capital cost and levelized cost of H₂ storage are analyzed.

The remaining of our paper is organized as follows: Section 2 discusses the methodology of estimating H₂ storage capacity and cost; section 3 presents the energy consumption in each state to determine the required storage capacity. Then, we summarize the H₂ storage capacity in potential geological sites. Finally, the cost of various geologic H₂ storage options is presented to assess the economic feasibility; section 4 reports this work's main conclusions and findings.

## 2. Methodology

This section discusses the assumptions and methods for estimating the capacity and cost of H₂ storage in depleted gas reservoirs, salt caverns, and saline aquifers. Based on these methods, we evaluate the potential H₂ storage capacity and cost in geological sites in the I-WEST region.

*2.1 Hydrogen storage capacity*

*2.1.1 Depleted gas reservoir*

In the estimation of H₂ storage capacity in depleted gas fields, several assumptions are made:
(1) The pressure and temperature gradients are 0.433 psi/ft and 15 °F/1000 ft, respectively (Lemieux et al. 2020);
(2) The cumulative production amount of natural gas under reservoir conditions is equal to the volume of stored H₂ in reservoir conditions;
(3) The cushion gas is H₂, and the volume percentage of cushion gas is 50% of the total storage volume (Lord 2009).

The methodology in this section is modified from the methods of H₂ storage estimation in the reference (Mouli-Castillo et al. 2021) by assuming the remaining gas is not considered as cushion gas. The natural gas cumulative production amount under standard conditions is obtained from the state databases. For the given average depth of a field, we can estimate the average pressure and temperature of the field with the pressure gradient (0.433 psi/ft) and temperature gradient (1.5 °F/100 ft) (Lemieux et al. 2020). With the average pressure and temperature of the field, we can calculate the compressibility factor ($z$) of natural gas using the Dranchuk and Abou-Kassem equation of state (DAK - EOS) (Dranchuk and Abou-Kassem 1975). The formation volume factor ($B_g$) can be calculated using the equation shown below:

$$B_g = \frac{V}{V_{sc}} = 0.0283 \frac{zT}{P} \qquad (1)$$

where the unit of $B_g$ is ft³/SCF, $P$ in psia, and $T$ in °R. Then, we can convert the natural gas cumulative production amount under standard conditions into reservoir conditions using $B_g$. Based on the assumption (2), we can obtain the volume of H$_2$ that can be stored underground. The underground H$_2$ storage volume can then be converted into the volume under standard conditions using equation (1). The $z$ factor of H$_2$ is obtained from NIST Reference Fluid Thermodynamic and Transport Properties Database (REFPROP) (Lemmon et al. 2007). Assuming the volume percentage of cushion gas (H$_2$) is 50%, we can determine the volume of cushion gas and working gas (50%). The workflow is shown in **Figure 4**.

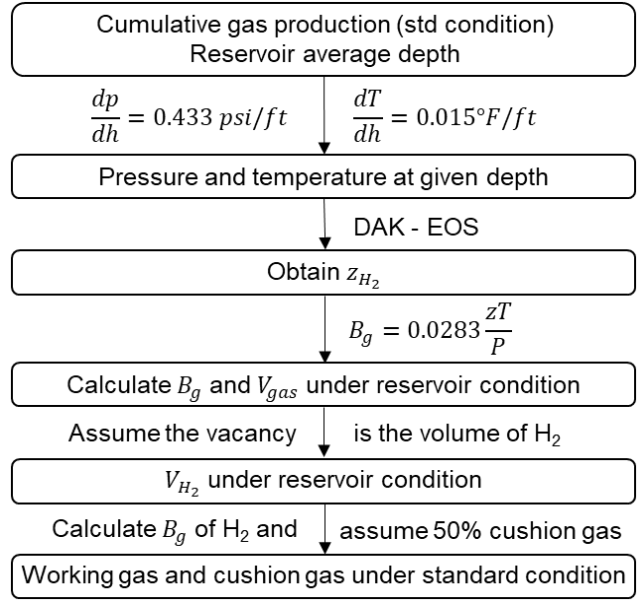

**Figure 4**. Workflow of the calculation of hydrogen storage volume in depleted gas reservoirs

*2.1.2 Salt cavern*

Several assumptions are made to model the salt caverns and estimate the volume of salt caverns based on the work of Lankof and Tarkowski (2020) and Pierce and Rich (1962):
(1) A salt cavern has a cylindrical shape with a specific diameter ($D$). The upper part of a cavern is a cone-shaped dome, with a height of 1/3 $D$. The lower part of a cavern is a conical incision, with a height of 1/6 $D$;
(2) Pressure gradient of fracture breakdown ($g_f$): 0.016 MPa/m;
(3) Minimum pressure gradient ($g_{min}$): 0.00835 MPa/m;
(4) Fraction of gas working capacity from the total volume: 80%;
(5) Temperature gradient: 0.027 °C/m.

The methodology of estimating single salt cavern storage capacity follows the work of Lankof and Tarkowski (Lankof and Tarkowski 2020). The volume of the cavity is calculated by assuming a cylindrical shape. The maximum and minimum pressure are computed using the equations shown below:

$$p_{max} = g_f h_n \qquad (2)$$
$$p_{min} = g_{min}(h_c - h_0) \qquad (3)$$

where $h_n$, $h_c$, and $h_0$ represent the depth to the top of the cavern neck, depth of the cavern center and depth of the cavern that can be emptied to zero pressure value. The values of $g_f$ and $g_{min}$ are 0.016 and 0.00835 MPa/m, respectively. The amount of $H_2$ stored in a single salt cavern at pressure $p$ is calculated by the equation:

$$m = f\frac{pV}{R^*Tz} \quad (4)$$

where $m$ is the amount of $H_2$ in the cavern, $f$ denotes the percentage of working gas, $T$ is the temperature, $R^*$ means the individual gas constant of $H_2$, which is 4121.73 $J/kg \cdot K$, and $z$ is the compressibility factor of $H_2$. The $z$ factor of $H_2$ is obtained from NIST REFPROP Database (Lemmon et al. 2007). The working capacity of $H_2$ can be obtained by calculating the difference of $m(p_{max})$ and $m(p_{min})$. The workflow is summarized in **Figure 5**.

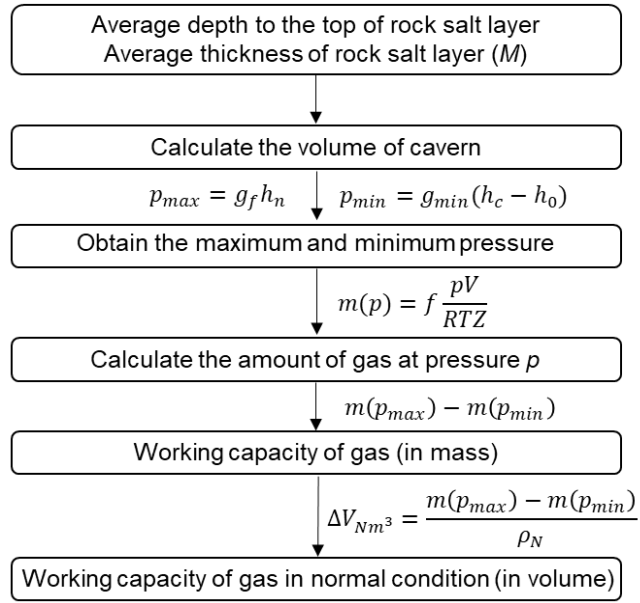

**Figure 5**. Workflow of the calculation of hydrogen storage volume in a single salt cavern.

The maximum number of salt caverns built in a specific region is estimated based on the potential area of rock salt layers. We assume that each salt cavern is built at the same depth with the same storage volume. To ensure the stability of the salt caverns, the distance between two adjacent salt caverns is four times the diameters of a single salt cavern (Bruno 2005). Therefore, we can estimate the maximum number of salt caverns using the total area of the rock salt layer divided by the square area occupied by a single salt cavern ($16D^2$).

*2.1.3 Saline aquifer*

Due to the similar characteristics of depleted gas reservoirs and saline aquifers, the assumptions and methodology of aquifer storage capacity estimation are similar to the ones of depleted gas reservoirs. One important parameter is the storage efficiency, which means the percentage of pore volume occupied by $H_2$. However, due to the limited research on $H_2$ storage in saline aquifers, the $H_2$ storage efficiency has not been analyzed. Based on the experimental and simulation data of $CO_2$ and natural gas storage in aquifer, the storage efficiency ranges from 5 to 20% (Tooseh et al. 2018; Bergmo et al. 2014), which depends on the aquifer properties. Herein, we assume the $H_2$ storage efficiency is 10%. The assumptions are shown below:
(1) The pressure and temperature gradients are 0.433 psi/ft and 15 °F/1000 ft, respectively (Lemieux 2020);
(2) The $H_2$ storage efficiency is 10%;
(3) The cushion gas is $H_2$, and the volume percentage of cushion gas is 80% (Lord 2009).

For a given depth of the aquifer, the average pressure and temperature of the aquifer can be estimated with the pressure and temperature gradients. With the assumption of 10% $H_2$ storage efficiency, the underground $H_2$ storage volume is obtained using the equation shown below:

$$V = Ah\varphi \cdot 10\% \tag{5}$$

where $A$ is the area of the aquifer, $h$ is the thickness of the aquifer, $\varphi$ represents the porosity of the aquifer. By calculating $B_g$ using equation (1), we can convert the underground $H_2$ storage volume to the $H_2$ storage volume under standard conditions. Assuming the cushion gas accounts for 80% of the total stored gas, the working gas capacity (20%) of aquifers can be determined. The workflow is shown in **Figure 6**.

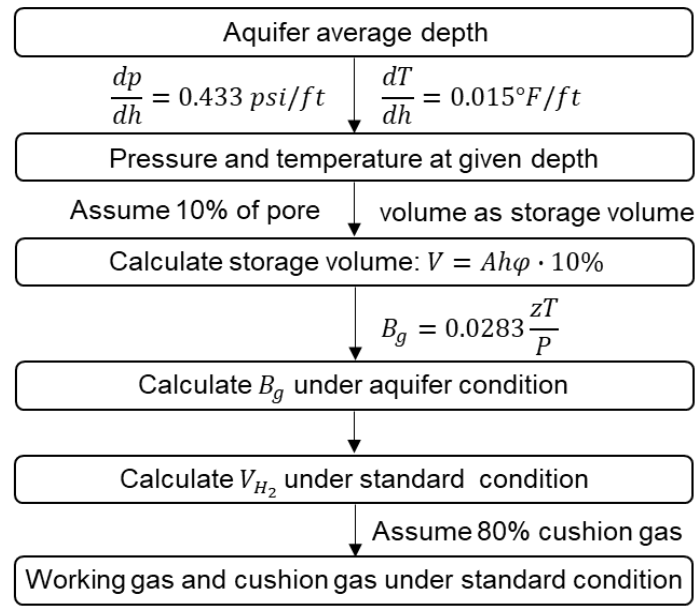

Figure 6. Workflow of the calculation of hydrogen storage volume in an aquifer.

*2.2 Hydrogen storage cost*

The economic feasibility of $H_2$ storage is generally based on the Hydrogen Geological Storage Model (H2GSM) proposed by Kobos et al. (2011). The cost estimation of each geological site includes two parts: the capital cost and the levelized cost of $H_2$ storage. The capital cost is a one-time expenditure, which includes the cost of well construction, equipment purchase, cushion gas, and potential site preparation. The levelized cost of $H_2$ storage estimates the average net present cost over its lifetime.

It includes the cost of equipment operation and maintenance, and resource consumption, together with the capital cost converted to each kilogram of $H_2$ over the lifetime.

*2.2.1 Capital cost*

The capital cost has four parts: 1) cushion gas cost; 2) geological site preparation cost; 3) compressor capital cost and 4) well capital cost. The detailed values are generally obtained from Lord et al. (2014) and summarized in **Table 1**.

Table 1. The capital cost of hydrogen storage

| Capital cost type | Name of capital cost | Depleted gas reservoir | Salt cavern | Saline aquifer |
|---|---|---|---|---|
| Cushion gas cost | Cushion gas percentage (%) | 50 | 30 | 80 |
|  | $H_2$ cost ($/kg $H_2$) | 5 | 5 | 5 |
| Geological site preparation cost | Mining cost ($/m$^3$) | 0 | 23 | 0 |
|  | Leaching plant cost ($/kg $H_2$) | 0 | 5 | 0 |
|  | Site characterization ($) | 0 | 115,000 | 10,300,000 |
|  | Mechanical integrity cost ($/kg) | 0 | 2.3 | 0 |
| Compressor capital cost | Total hours of operation (hour/year) | 5,600 | 5,600 | 5,600 |
|  | Compressor size ($H_2$ kg/hour) | 2,000 | 2,000 | 2,000 |
|  | Compressor capacity (kton $H_2$) | 11.2 | 11.2 | 11.2 |
|  | Capital cost per compressor ($) | 9,179,700 | 9,179,700 | 9,179,700 |
|  | Compressor power (kWh/kg $H_2$) | 2.2 | 2.2 | 2.2 |
|  | Cost of electricity ($/kWh) | 0.128 | 0.128 | 0.128 |
|  | Water requirement (L/kg $H_2$) | 50 | 50 | 50 |
|  | Water & cooling cost ($/100L $H_2O$) | 0.02 | 0.02 | 0.02 |
| Well capital cost | Well cost ($/per well) | 260,000 | 1,150,000 | 1,150,000 |

The cost of $H_2$ production keeps decreasing with the development of technology. The European Commission's July 2020 $H_2$ strategy shows that the green $H_2$ production cost is between $3/kg and $6.55/kg (van Renssen 2020). Herein, the $H_2$ cost is considered as $5/kg.

The cost of geological site preparation is highly dependent on the type of geological site. Both depleted gas reservoirs and saline aquifers have natural porous media to store $H_2$. Due to the previous exploration and development experience, no extra effort is needed for depleted gas reservoirs. However, further analysis is required for saline aquifers to better understand the geological structures to ensure a domal shape storage space and an impermeable rock on the top. For the salt cavern, the main capital cost is related to cavity erection and geological survey, which includes the mining cost, the leaching plant cost, site characterization cost, and mechanical integrity test cost.

The capital cost of a compressor includes two parts: the purchase of compressors and the cost of compressing cushion gas. We assume that two-thirds of the year is considered for injection while one-third of the year is used for extraction (Lord 2009).

Therefore, the total hours of operation are 5,600 hours. The compressor size is assumed to be 2,000 kg/hr, which means that one compressor can compress 2,000 kg of $H_2$ to given pressure in one hour. The required compressors can be calculated based on the working gas capacity (WGC). The cost of cushion gas compression involves electricity cost (EC) and water and cooling cost (WCC), which can be calculated by the equations below:

$$EC = WGC \times CP \times UPE \qquad (6)$$
$$WCC = WGC \times WR \times UPWC \qquad (7)$$

where CP represents compressor power (kWh/kg $H_2$), UPE denotes the unit price of electricity ($/kWh), WR is water requirement (L/kg $H_2$), and UPWC means the unit price of water and cooling ($/100L $H_2O$).

For depleted gas reservoirs and aquifers, we assume each well is in charge of 3,000 tons of $H_2$ injection. The number of wells can be calculated based on the $H_2$ storage amount. For salt caverns, we assume one well is drilled for one salt cavern. For all three geological sites, the same well is used for both $H_2$ injection and extraction. Compared with other geological sites, the cost of wells in depleted gas reservoirs is lower because the previously drilled wells may be reused for $H_2$ injection and extraction after some repairing procedures.

The total capital cost (TCC) is the sum of cushion gas cost, geological site preparation cost, compressor capital cost, and well capital cost.

## 2.2.2 Levelized cost

The levelized cost of $H_2$ storage consists of three main parts: 1) levelized total capital cost; 2) compressor operation and maintenance cost (COMC); and 3) well operation and maintenance cost (WOMC). The detailed values are generally obtained from Lord et al. (Lord et al. 2014) and summarized in **Table 2**. The values of three different geological sites are similar. The only difference is the levelized $H_2$ well cost due to the previously erected infrastructure in the depleted gas reservoir.

Table 2. Levelized cost of hydrogen storage

| Levelized cost type | Name of levelized cost | Depleted gas reservoir | Salt cavern | Salien aquifer |
|---|---|---|---|---|
| Levelized total capital cost | Discount rate | 0.1 | 0.1 | 0.1 |
| | Well lifetime (year) | 40 | 40 | 40 |
| | Capacity factor | 0.8 | 0.8 | 0.8 |
| Compressor operation and maintenance cost | Electricity cost ($/kg $H_2$) | 0.2816 | 0.2816 | 0.2816 |
| | Water and cooling cost ($/kg $H_2$) | 0.01 | 0.01 | 0.01 |
| Well operation and maintenance cost | $H_2$ well cost ($/ kg $H_2$) | 0.0105 | 0.04627 | 0.04627 |
| | $H_2$ surface pipeline cost ($/ kg $H_2$) | 0.00403 | 0.00403 | 0.00403 |

The equation of levelized total capital cost (LTCC) is shown below:

$$LTCC = (TCC \times CRF)/CF \qquad (8)$$

TCC is calculated as the sum of all capital costs. The capacity factor (CF) is assumed to be 0.8. The capital recovery factor (CRF) is obtained with the equation shown below:

$$CRF = \frac{r(1+r)^t}{(1+r)^t-1} \quad (9)$$

where *r* denotes the discount rate, and *t* represents the economic lifetime. The levelized cost of hydrogen storage (LCHS) is calculated based on the equation shown below:

$$LCHS = \frac{LTCC}{WGC} + COMC + WOMC \quad (10)$$

## 3. Results and analysis

In this section, the energy demand in the I-WEST region is estimated based on the population and energy consumption per capita. The capacity and cost of $H_2$ storage are calculated based on the aforementioned methods. Finally, the energy demand, $H_2$ storage capacity and cost are summarized by states.

*3.1 Hydrogen storage capacity*

In this subsection, we will discuss the $H_2$ storage capacity in depleted gas reservoirs, salt carverns, and saline aquifers.

*3.1.1 Depleted gas reservoirs*

We selected depleted gas reservoirs with high cumulative production in the I-WEST region to ensure the large storage capacity. We calculate the working gas capacity of $H_2$ storage based on the methodology mentioned in Section 2.1.1. The results are presented in **Figure 7**. The sizes of dots represent the working gas capacity of depleted gas reservoirs, while the colors of dots represent the major types of formations in depleted gas reservoirs. We can observe that the number of depleted gas fields with working gas capacity less than 200 kton, between 200 and 600 kton, and higher than 600 kton is 14, 9, and 4, respectively. In our calculation, we consider the $H_2$ as cushion gas to ensure the high purity of produced $H_2$. If the cushion gas is the remaining natural gas in the field, the working gas capacity of $H_2$ storage should be much higher.

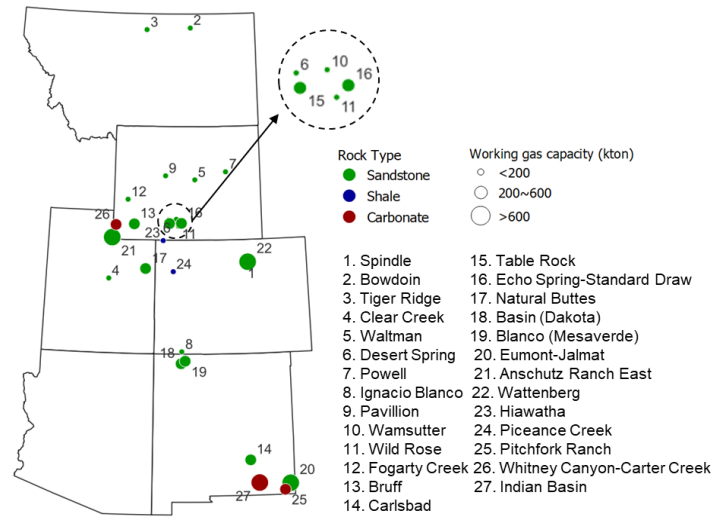

**Figure 7**. Working gas capacity of $H_2$ storage in depleted gas reservoirs in the I-WEST region. The labels in the figure represent the names of fields. The name in the parenthesis is the formation of the field.

*3.1.2 Salt cavern*

Firstly, the sites with large-scale rock salt deposits in the I-WEST region are considered the potential locations for the erection of salt caverns. Then, the selection of ideal locations for salt caverns includes the top depth of the rock salt layer and the thickness of the rock salt layers. Considering the stability of salt caverns and their possible effect on underground water, the suitable top depth of the rock salt layer ranges from 500 to 1,800 m (Lankof and Tarkowski 2020; Pierce and Rich 1962). The minimum thickness of the rock salt layer is considered to be 122 m (400 ft) to ensure a large storage capacity and cost-effectiveness.

Following the aforementioned methods, we select ten potential locations to build salt caverns for $H_2$ storage. The results are shown in **Figure 8**. The sizes of dots represent the working gas capacity of a single salt cavern, while the colors of dots represent the type of salt caverns. We can observe that a single salt cavern has a storage capacity of several thousand tons. Most of the salt caverns are built in Arizona, Colorado and Utah. Since Arizona does not have any depleted gas reservoirs for $H_2$ storage as shown in **Figure 7**, salt caverns can be the alternative geologic sites for $H_2$ storage in Arizona.

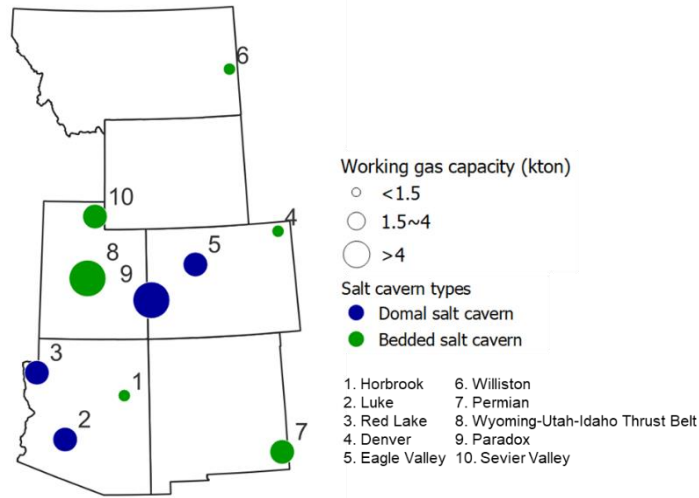

**Figure 8**. Working gas capacity of $H_2$ storage in a single salt cavern in the I-WEST region. The labels in the figure represent the names of basins.

The total working capacity of salt caverns in the desired region is summarized in **Figure 9**. The total working capacity is calculated based on the working capacity of a single salt cavern and the potential number of salt caverns built in that region. Based on our estimation, the number of salt caverns in a basin ranges from several hundreds to several thousands. The total working gas capaicity of salt caverns is at the scale of million tons. The sizes of dots represent the total working gas capacity of salt caverns, while the colors of dots represent the type of salt caverns.

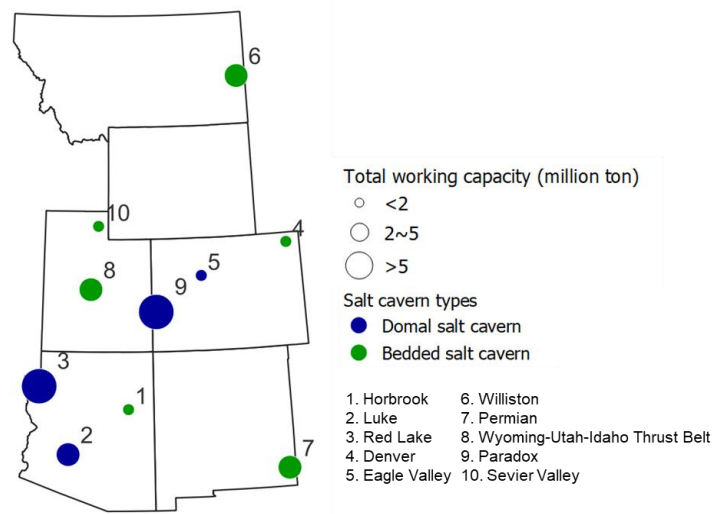

**Figure 9**. Total working gas capacity of hydrogen storage in salt caverns in the I-WEST region. The labels in the figure represent the names of basins.

*3.1.3 Saline aquifer*

Generally, an aquifer suitable for storage has a similar geological requirement as a depleted gas reservoir, including the high porosity and permeability of porous media with impermeable cap rocks overlaid (Sainz-Garcia et al. 2017). The aquifers with large drainage areas are selected to ensure a large storage volume. Additionally, high porosity and permeability aquifers are preferred due to their better storage capability and deliverability. Following these criteria, 12 saline aquifers are selected as the potential storage sites in the I-WEST region.

Based on the aforementioned methods, the working gas capacity of $H_2$ storage in aquifers is shown in **Figure 10**. The sizes of dots represent the working gas capacity of aquifers, while the colors of dots represent the lithology of aquifers. The storage capacity of aquifers includes several formations in the basins. We can observe that the working gas capacity in many regions is more than 1,000 million tons due to the wide distribution of underground water. However, site characterization is necessary to narrow down the potential area and determine the final storage sites to ensure the sealing strength.

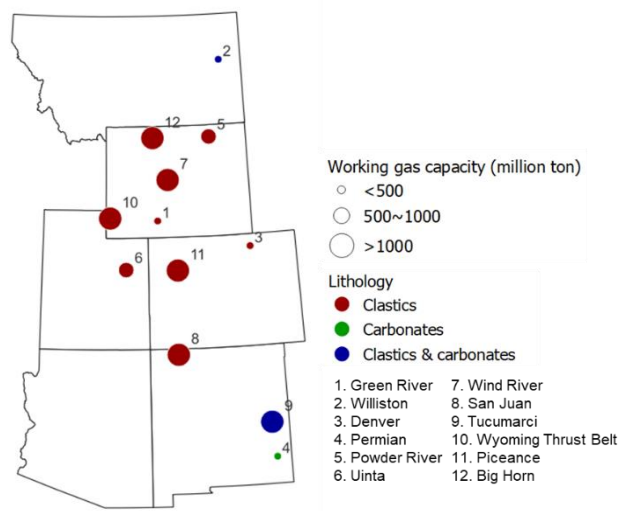

**Figure 10**. Working gas capacity of $H_2$ storage in aquifers in the I-WEST region. The labels are the names of the basins where the aquifers are located in.

To give a rough estimation of the storage capacity of an aquifer site, the Baker dome in Four Corners Platform, San Juan Basin, is selected as the target $H_2$ storage site (Kelley 1957). The tectonic trap of the dome allows the formation of a gas cap, which contributes to the recovery of $H_2$ (Foh et al. 1979). The detailed information (Fassett 1991) on $H_2$ storage in the Baker dome is shown in **Table 3**.

*3.2 Energy demand*

We estimate the total amount of energy required per year in each state based on the data from EIA (EIA Independent Statistics and Analysis). Considering the energy value of $H_2$ is 3 kWh/m$^3$ (Lankof and Tarkowski 2020), we converted the energy consumption in the I-WEST region to the amount of $H_2$ required, as shown in **Figure 11**. We can observe that Arizona and Colorado have higher energy demands than other states in the I-WEST region. The major energy consumption sectors in the I-WEST region are the industrial and transporation sectors.

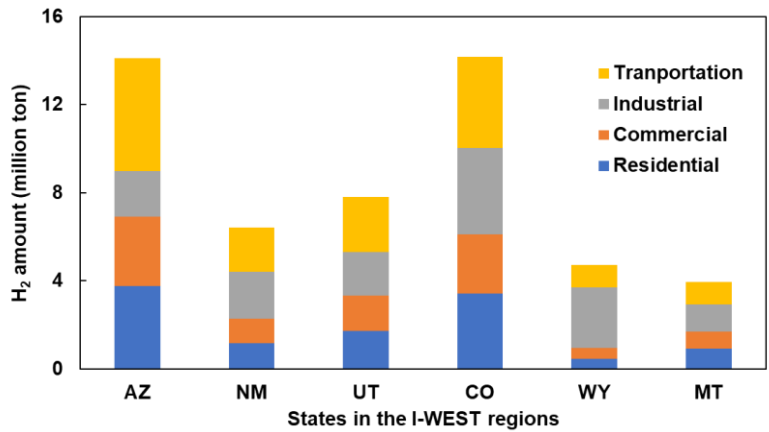

**Figure 11**. Energy consumption per year in I-WEST region.

To analyze if the $H_2$ storage capacity meets the energy demand in the I-WEST region, we summarize each state's energy demand and $H_2$ storage capacity in **Figure 12**. It is worth noting that, due to the limited site characterization and high cost of aquifer storage, only the $H_2$ storage capacities in depleted gas reservoirs and salt caverns are considered. Comparing the energy demand and storage capacity, the $H_2$ storage capacities in depleted gas reservoirs and salt caverns can meet 72% of the total energy demand in the I-WEST region in 2020. According to the International Energy Agency (IEA) prediction, $H_2$ will account for about 10% of total energy consumption in 2050 (Bouckaert et al. 2021). **Figure 12** suggests that $H_2$ storage capacity of the depleted gas reservoirs and salt caverns meet at least 30% of energy demand in each state in the I-WEST region, which is higher than the percentage of $H_2$ in energy consumption predicted by IEA. However, the estimation is made at the state level. Considering the different energy consumptions and geologic structures in a city, detailed plans should be made. For example, for the regions away from depleted gas reservoirs and unsuitable for the erection of salt caverns, an aquifer (if available nearby) should be considered as the first choice of the $H_2$ storage site.

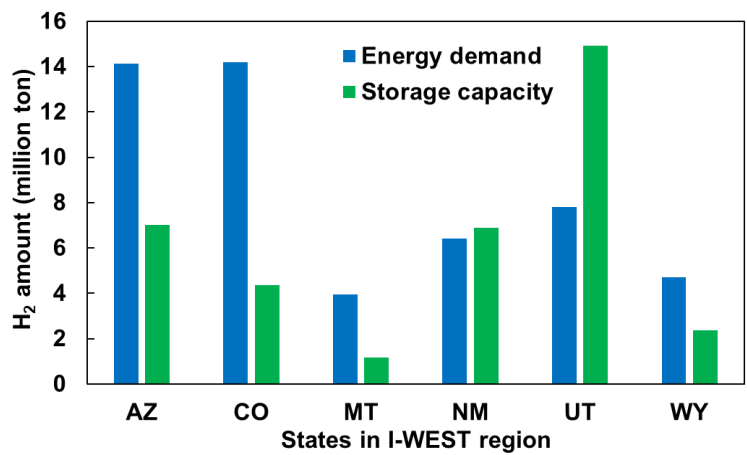

**Figure 12**. Energy demand and hydrogen storage capacity in the I-WEST regions in 2020. The hydrogen storage capacity only includes the storage capacities in depleted gas reservoirs and salt caverns. AZ:

Arizona, CO: Colorado, MT: Montana, NM: New Mexico, UT: Utah, WY: Wyoming.

*3.3 Hydrogen storage cost*

We calculated the capital and levelized cost of $H_2$ storage in three typical case studies for depleted gas reservoirs, salt caverns and saline aquifers. The characterizations of three geologic sites are shown in **Table 3**.

Table 3. Geological site characterization

|  | Depleted gas reservoir | Salt deposits per cavern | Aquifer |
|---|---|---|---|
| Geological site | Watternberg field (CO) | Red Lake (AZ) | Baker dome (CO) |
| Storage volume underground (million ft$^3$) | 8,200 | 15.5 | 5,602 |
| Average depth (ft) | 8,000 | 4,000 | 4,717 |
| Average Pressure (psi) | 3,479 | 1,732 | 2,057 |
| Average temperature (F) | 180 | 122 | 128 |
| Total $H_2$ storage amount (kton) | 3,546 | 4.2 | 1613 |
| Working gas percentage (%) | 50 | 80 | 20 |
| Working gas capacity (kton) | 1,773 | 3.4 | 323 |
| Cushion gas amount (kton) | 1,773 | 0.8 | 1,290 |

The capital and levelized cost of $H_2$ storage of the three geologic sites are shown in **Tables 4** and **5**. The proportion of each cost type is shown in **Figures 13** and **14**. As shown in **Figure 13**, the cushion gas cost accounts for more than 80% of the capital cost of depleted gas reservoir or saline aquifer, while the geologic site preparation cost is the major cost of $H_2$ storage in salt cavern. For the levelized cost of $H_2$ storage, the levelized total capital cost is the major part regardless of the type of geologic site, which is due to the high value of total capital cost. For salt caverns, the total capital cost is mainly affected by the mining and leaching cost, which is determined by the cavity volume. However, for depleted gas reservoirs or aquifers, the total capital cost can be significantly reduced if we lower the cushion gas cost, which is possible by changing the type of cushion gas.

Table 4. The capital cost of hydrogen storage of three geological sites

|  | Depleted gas reservoir | Salt cavern | Saline aquifer |
|---|---|---|---|
| Geological site | Watternberg field (CO) | Red Lake (AZ) | Baker dome (CO) |
| Cushion gas cost (million $) | 8,864 | 4.2 | 6,452 |
| Geologic site preparation cost (million $) | 0 | 36.7 | 10 |
| Compressor capital cost (million $) | 1,977 | 9.4 | 642 |
| Well capital cost (million $) | 154 | 1.2 | 124 |
| Total capital cost (million $) | 10,994 | 51 | 7,229 |

Table 5. Levelized cost of hydrogen storage of three geologic sites

| | Depleted gas reservoir | Salt cavern | Saline aquifer |
|---|---|---|---|
| Geologic site | Watternberg field (CO) | Red Lake (AZ) | Baker dome (CO) |
| Levelized total capital cost ($/kg) | 0.7927 | 1.9552 | 2.8644 |
| Compressor levelized cost ($/kg) | 0.2916 | 0.2916 | 0.2916 |
| Well and surface pipeline levelized cost ($/kg) | 0.0146 | 0.0503 | 0.0503 |
| Levelized cost of $H_2$ storage ($/kg) | 1.0989 | 2.2971 | 3.2063 |

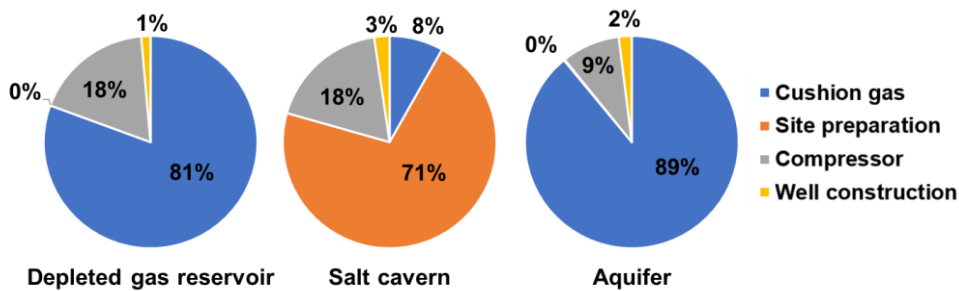

**Figure 13**. Pie charts of the capital cost of $H_2$ storage in the depleted gas reservoir, salt cavern, and saline aquifer.

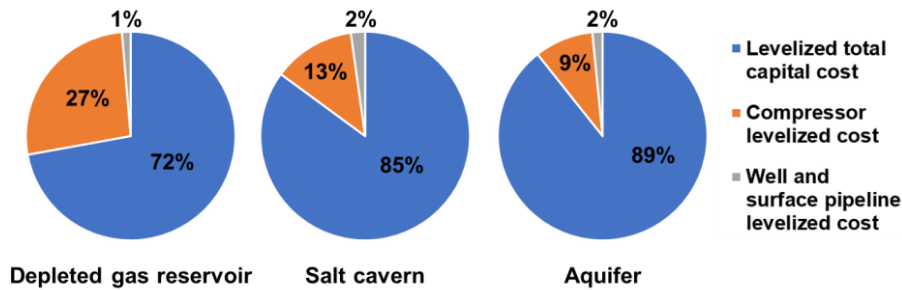

**Figure 14**. Pie charts of levelized cost of hydrogen storage in the depleted gas reservoir, salt cavern, and aquifer.

To evaluate the effect of cushion gas types on $H_2$ storage cost, we consider natural gas and nitrogen as alternative cushion gases. The comparisons are summarized in **Figures 15** and **16**. Due to the contamination of cushion gas, the extra cost should be considered in $H_2$ purification. According to previous analysis (He 2017; Nordio et al. 2021), the cost of $H_2$ purification ranges from $1 to $8.3/kg $H_2$, which depends on the initial $H_2$ percentage, target $H_2$ purity, and types of mixed gases. Herein, we consider the purification cost as $2/kg $H_2$. We can observe that the total capital cost reduces by about 8% and 35% using natural gas or $N_2$ as cushion gas in depleted gas reservoirs and aquifers, respectively. The reduction is more significant for saline aquifers storage. This

is due to the high percentage of cushion gas requirement. In addition, we calculate the threshold of purification cost, which refers to the cost that does not contribute to the change of cushion gas type. For the depleted gas reservoirs, the purification cost thresholds of natural gas and $N_2$ are \$2.50 and \$2.55/kg $H_2$, respectively. For the saline aquifers, the purification cost thresholds of natural gas and $N_2$ are \$9.83 and \$9.86/kg $H_2$, respectively, which are higher than the current $H_2$ purification cost. The high value of the purification cost threshold in saline aquifers indicates that it is economical to use another type of gas as cushion gas for $H_2$ storage in saline aquifers.

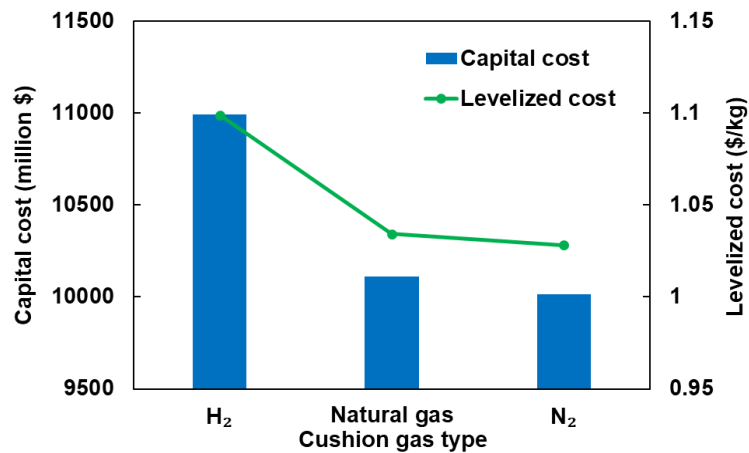

**Figure 15**. Capital cost and levelized cost of $H_2$ storage in depleted gas reservoirs with different types of cushion gas.

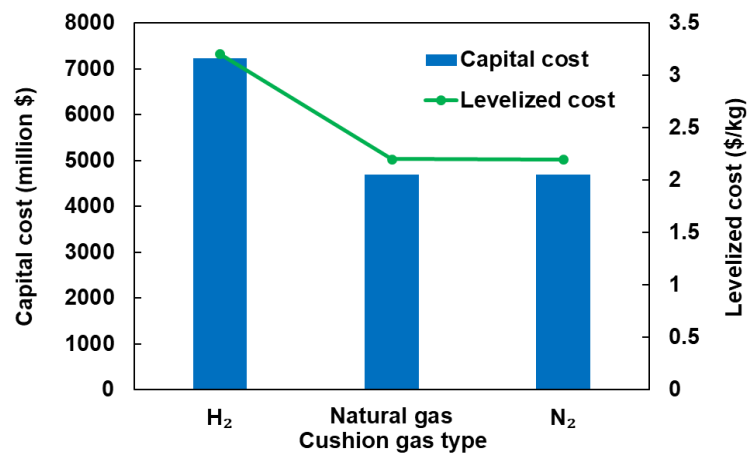

**Figure 16**. Capital cost and levelized cost of $H_2$ storage in saline aquifers with different types of cushion gas.

The impact of working gas capacity on storage cost is analyzed in **Figures 17**. We assume the average depth of the storage site is 1,000 m. The figure shows that $H_2$ storage in saline aquifers always has the highest cost regardless of the working gas capacity due to the large requirement for cushion gas. When the working gas capacity is higher than 0.25 kton, $H_2$ storage in the depleted gas reservoir is the most cost-effective choice. The reason is that the high working gas capacity requires a large storage volume, which leads to a high mining and leaching cost for $H_2$ storage in salt caverns. The levelized cost of $H_2$ storage first decreases sharply with the increased working gas capacity. Then the curves become flat, indicating that the levelized cost is not significantly affected by the working gas capacity. The levelized cost of $H_2$ storage in the depleted gas reservoir, salt cavern, and aquifer at high working gas capacity (100 kton) is about $1.3, $2.3, and $3.4/kg $H_2$, respectively. From equation (10), the working gas capacity primarily affects the first term of levelized cost of $H_2$ storage (LCHS). For the fixed capital cost which is not calculated in the unit price ($/kg), the increase in storage capacity will lead to a decrease in the first term of LCHS and therefore reduces the value of LCHS. With the continuous growth of working gas capacity, the fixed capital cost becomes relatively low compared with the capital cost that is proportional to the working gas capacity. Thus, an increasing linear trend exists in capital cost curves while the levelized cost curves become flat at high working gas capacity.

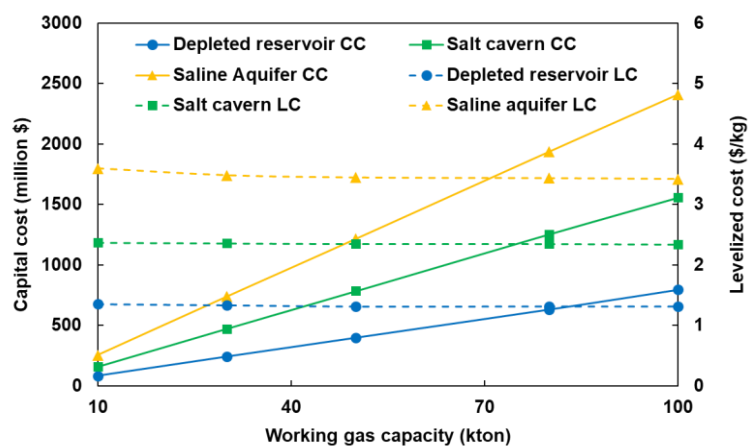

**Figure 17**. The total capital cost and levelized cost of $H_2$ storage with different working gas capacities. In the legend, "CC" means capital cost while "LC" represents levelized cost.

Based on IEA prediction, $H_2$ will account for about 10% of total energy consumption in 2050 (Bouckaert et al. 2021). Therefore, the capital cost of 10% of energy demand and levelized cost of $H_2$ storage are estimated in **Figure 18**. The high capital cost is caused by the large storage capacity whereas the high levelized cost is due to the high percentage of $H_2$ storage in salt caverns.

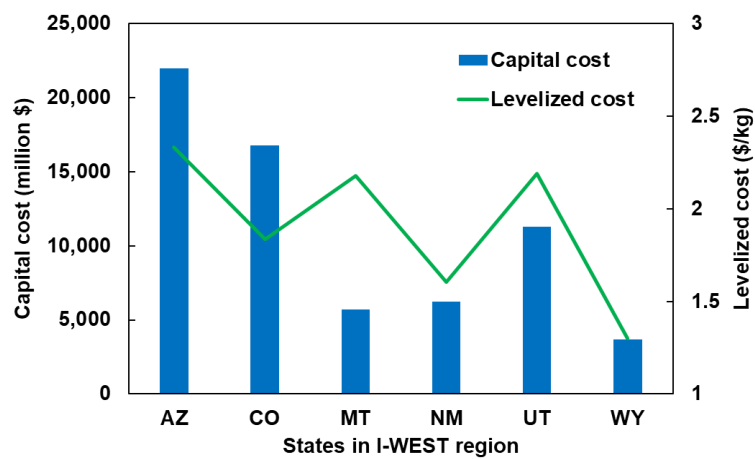

**Figure 18**. Capital cost and levelized cost of $H_2$ storage in the I-WEST regions. AZ: Arizona, CO: Colorado, MT: Montana, NM: New Mexico, UT: Utah, WY: Wyoming.

## 4. Conclusion

We analyzed the technical and economic feasibility of $H_2$ geologic storage in the I-WEST region. We found that $H_2$ geologic storage can meet at least 30% of total energy consumption in each state of the I-WEST region. The levelized cost ranges from $1.3 to $2.3/kg $H_2$ in the states considered. The main findings are summarized as follows:

(1) We developed workflows to estimate the $H_2$ storage capacity of the depleted gas reservoirs, salt caverns, and saline aquifers. The suitable geologic sites for $H_2$ storage are identified, and the working gas capacity is estimated.

(2) The capital costs of $H_2$ storage in depleted gas reservoirs, salt caverns and saline aquifers are estimated at 10,994, 51 and 7,229 million dollars with the working gas capacities of 1,773, 3.4 and 323 ktons, respectively. The cushion gas cost accounts for more than 80% of the capital cost of depleted gas reservoirs or saline aquifers, while the geologic site preparation cost is the major cost for $H_2$ storage in salt caverns.

(3) Due to the high cost of $H_2$, we evaluated the economic feasibility of using natural gas and nitrogen as alternative cushion gases. Assuming the purification cost of $H_2$ is $2/kg, the total capital cost reduces by about 8% and 35% by using natural gas or nitrogen as cushion gas in depleted gas reservoirs and aquifers, respectively. Therefore, further analysis is needed to optimize the selection of the cushion gas in aquifer storage.

(4) We analyzed the effect of working gas capacity on storage cost. $H_2$ storage in aquifers has the highest cost regardless of the working gas capacity. On the other hand, $H_2$ storage in a depleted gas reservoirs is the most cost-effective option. The levelized cost of $H_2$ storage in a depleted gas reservoir, salt cavern, and aquifer with large-scale storage capacity is about $1.3, $2.3, and $3.4/kg $H_2$, respectively.

## Nomenclature

| | | |
|---|---|---|
| $COMC$ | = | Compressor operation and maintenance cost |
| $CF$ | = | Capital factor |
| $CP$ | = | Compressor power |
| $CRF$ | = | Capital recovery factor |
| $EC$ | = | Electricity cost |
| $LCHS$ | = | Levelized cost of hydrogen storage |
| $LTCC$ | = | Levelized total capital cost |
| $TCC$ | = | Total capital cost |
| $UPE$ | = | Unit price of electricity |
| $UPWC$ | = | Unit price of water and cooling |
| $WCC$ | = | Water and cooling cost |
| $WGC$ | = | Working gas capacity |
| $WOMC$ | = | Well operation and maintenance cost |
| $WR$ | = | Water requirement |
| $A$ | = | Area of aquifer |
| $B_g$ | = | Formation volume factor |
| $D$ | = | Diameter of cavern |
| $f$ | = | Percentage of working gas |
| $g_f$ | = | Fracture breakdown pressure gradient |
| $g_{min}$ | = | Minimum pressure gradient |
| $h_0$ | = | Depth of the cavern with zero value of Pressure |
| $h_c$ | = | Depth of the cavern center |
| $h_n$ | = | Depth to the top of the cavern neck |
| $m$ | = | Hydrogen amount in the cavern |
| $p$ | = | Pressure |
| $r$ | = | Discount rate |
| $R^*$ | = | Individual gas constant of hydrogen |
| $t$ | = | Economic lifetime |
| $T$ | = | Temperature |
| $V$ | = | Volume |
| $z$ | = | Compressibility factor |
| $\varphi$ | = | Porosity |

## Acknowledgment

This work was supported by Los Alamos National Laboratory Technology Evaluation & Demonstration funds.